\documentclass[RNAAS]{aastex62}

\submitjournal{RNAAS}

\usepackage{natbib}
\usepackage{url}

\newcommand{\as}{\ensuremath{^{\prime\prime}}} 

\begin{document}
	
\title{Evidence for unresolved exoplanet-hosting binaries in Gaia DR2}

\correspondingauthor{Daniel F. Evans}
\email{d.f.evans@keele.ac.uk}

\author[0000-0002-5383-0919]{Daniel F. Evans}
\affil{Astrophysics Group, Keele University, \\
	Staffordshire, ST5 5BG, UK}


\section{}

This note describes an effort to detect additional stellar sources in known transiting exoplanet (TEP) systems, which are unresolved or barely resolved in the Gaia Data Release 2 (DR2) catalogue \citep{2016A+A...595A...1G, 2018arXiv180409365G}. The presence of multiple unresolved stars in photometric and spectroscopic observations of a transiting planetary system biases measurements of the planet's radius, mass, and atmospheric conditions (e.g. \citealt{2011ApJS..197....3B}, \citealt{2016MNRAS.463...37S}, \citealt{2016ApJ...833L..19E}). In addition to the effect on individual planetary systems, the presence of unresolved stars across the sample of known exoplanets biases our overall understanding of planetary systems, due to the systematic underestimation of both masses and radii \citep{2015ApJ...805...16C}.

Sources with separations below approximately 100\,mas are unresolved by the Gaia satellite. At wider separations, the probability that Gaia will determine the presence of two is dependent on the separation and flux ratio of the pair. During the data processing, detected companion stars may be rejected as a duplicate of the brighter star. \citep{2018arXiv180409366L}. Whilst sources for which duplicates were identified are logged in Gaia DR2, this is not a reliable indicator of binarity, as many spurious duplicates are correctly removed during the processing. Instead, this work uses the Astrometric Goodness of Fit in the Along-Scan direction (\texttt{GOF\_AL}) and the Astrometric Excess Noise as indicators of poorly-resolved binaries. The former is the `gaussianised Chi-square' for the astrometric fit, with mean zero and standard deviation one, and values greater than 3 indicating poor agreement between the model and data. The astrometric excess noise is an additional uncertainty assigned to measurements of a source to encompass modelling errors. The significance of the astrometric excess noise, \texttt{D}, is also computed and available in the DR2 database. Half of sources are expected to have a \texttt{D} value of zero, while the remaining half are distributed in the positive part of a normal distribution with mean zero and standard deviation one.

A list of coordinates of known TEP systems was drawn from the the \texttt{TEPCat}\footnote{\url{http://www.astro.keele.ac.uk/jkt/tepcat/}} catalogue \citep{2011MNRAS.417.2166S}, including both the well-studied and little-studied systems. These were cross-matched with the Gaia DR2 catalogue by angular separation (accounting for proper motion) and brightness (comparing V magnitudes to Gaia G magnitudes). Many known close binaries in the exoplanet host star sample have highly significant \texttt{GOF\_AL} and \texttt{D} values, such as WASP-20AB with \texttt{D}\,$=4720$ and \texttt{GOF\_AL}\,$=124$. As the cutoff for \texttt{D} and \texttt{GOF\_AL} is reduced, an increasing proportion of systems have high resolution imaging or interferometric observations that rule out most possible binary companions. The final cutoff was set manually at \texttt{D}\,$>5$ and \texttt{GOF\_AL}\,$>20$ to best match the boundary between confirmed binaries and confirmed singles.

Systems matching both of these criteria are listed in Table~\ref{tab:objects}, grouped into objects for which a sufficiently close binary companion is known, typically through adaptive optics imaging or interferometry, and those objects with no such companion detected. Stars that are exceptionally bright or have high proper motion are highlighted, as it is plausible that the large offset is related to difficulties in modelling saturated or fast-moving stars, rather than unresolved binarity.

\begin{table}
	\centering
	\caption{Transiting exoplanet systems in with evidence for unresolved binarity in Gaia DR2. \label{tab:objects}}
	\begin{tabular}{cccl}
		\tablewidth{0pt}
		\hline
		\hline
		System & \texttt{D} & \texttt{GOF\_AL} & Comments \\
		\hline
		\multicolumn{4}{l}{Systems with confirmed close companions}\\
		\decimals
		EPIC 246067459 &   345.0 &  67.1 & Companion at 0.35\as\ \citep{2018arXiv180107959S} \\
		Kepler-13      &    55.0 &  25.6 & Bright companion at 1.1\as\ \citep{2014ApJ...788...92S} \\
		Kepler-64      &    98.5 &  50.7 & Eclipsing binary with companion at 0.7\as\ \citep{2013ApJ...768..127S} \\
		Kepler-108     &  5530.0 & 214.7 & Bright companion at 1.1\as\ \citep{2014ApJ...791...35L} \\
		Kepler-132     &  1180.0 & 148.3 & Companion at 0.87\as\ \citep{2016AJ....152....8K} \\
		Kepler-133     &    24.7 &  31.4 & Companion at 0.07\as\ \citep{2016AJ....152....8K} \\
		Kepler-296     &  3160.0 & 179.1 & Companion at 0.21\as\ \citep{2016AJ....152....8K} \\
		Kepler-326     &    35.4 &  25.0 & Companion at 0.05\as\ \citep{2016AJ....152....8K} \\
		Kepler-336     &  3720.0 & 230.3 & Companion at 0.27\as\ \citep{2014ApJ...791...35L} \\
		Kepler-345     &   303.0 &  81.7 & Companion at 0.08\as\ \citep{2016AJ....152....8K} \\
		Kepler-365     &    30.4 &  24.3 & Companion at 0.77\as\ \citep{2017AJ....153...66Z} \\
		Kepler-369     &  3010.0 & 185.5 & Companion at 0.13\as\ \citep{2016AJ....152....8K} \\
		Kepler-383     &  4600.0 & 188.1 & Companion at 0.31\as\ \citep{2014ApJ...791...35L} \\
		Kepler-400     &  5290.0 & 204.3 & Companion at 0.52\as\ \citep{2016AJ....152...18B} \\
		Kepler-437     & 44900.0 & 495.0 & Companion at 0.18\as\ \citep{2016AJ....152....8K} \\
		Kepler-438     &   139.0 &  41.9 & Companion at 0.44\as\ \citep{2016AJ....152....8K} \\
		LHS 6343       & 23500.0 & 462.5 & Companion at 0.7\as\ \citep{2011ApJ...730...79J} \\
		NLTT 41135     &    78.9 &  37.5 & Brighter companion at 2.4\as\ \citep{ 	2011ApJ...730...79J} \\
		WASP-2         &    11.7 &  23.7 & Companion at 0.7\as\ \citep{2007MNRAS.375..951C} \\
		WASP-11        &    75.4 &  37.0 & Companion at 0.36\as\ \citep{2015ApJ...800..138N} \\
		WASP-20        &  4720.0 & 124.2 & Companion at 0.26\as\ \citep{2016ApJ...833L..19E} \\
		WASP-72        &     7.4 &  21.6 & Companion at 0.64\as\ (Evans et al., in prep.) \\
		WASP-76        &   130.0 &  58.2 & Companion at 0.43\as\ \citep{2015A+A...579A.129W} \\
		WASP-103       &   291.0 & 106.4 & Companion at 0.24\as\ \citep{2015A+A...579A.129W} \\
		WTS-2          &    48.9 &  35.7 & Companion at 0.6\as\ \citep{2014MNRAS.440.1470B} \\
		\hline
		\multicolumn{4}{l}{Systems that may contain unresolved companion stars}\\
		55 Cnc         &    14.4 &  28.8 & Bright, high proper motion \\
		CoRoT-26       &   155.0 &  45.6 & \\
		GJ 1214        &    35.2 &  24.1 & High proper motion \\
		GJ 3470        &    12.4 &  23.8 & High proper motion \\
		HATS-12        &   327.0 &  79.5 & \\
		HATS-39        &    61.7 &  49.2 & \\
		Kepler-84      &  6560.0 & 244.1 & No companion in Robo-AO \citep{2014ApJ...791...35L} \\
		Kepler-241     &    69.9 &  29.8 & No companion in Robo-AO \\
		Kepler-313     &   114.0 &  49.7 & No companion in Robo-AO \\
		Kepler-362     &    27.8 &  21.9 & No companion in Robo-AO \\
		Kepler-388     &   354.0 &  81.9 & No companion in Robo-AO \\
		Kepler-399     &    86.6 &  36.4 & No companion within 3\as\ in Robo-AO \\
		Kepler-420     &  1060.0 & 118.0 & No companion in Robo-AO \\
		Kepler-449     & 31500.0 & 565.9 & No companion in Robo-AO \\
		Qatar-3        &    13.0 &  23.8 & \\
		WASP-22        &    12.5 &  33.1 & Long period radial velocity signal \citep{2010AJ....140.2007M} \\
		WASP-66        &    11.5 &  23.2 & \\
		WASP-105       &   648.0 & 148.1 & \\
		WASP-107       &     8.8 &  22.7 & \\
		\hline
	\end{tabular}
\end{table}

\acknowledgments

This work has made use of data from the European Space Agency (ESA) mission {\it Gaia} (\url{https://www.cosmos.esa.int/gaia}), processed by the {\it Gaia} Data Processing and Analysis Consortium (DPAC, \url{https://www.cosmos.esa.int/web/gaia/dpac/consortium}). Funding for the DPAC has been provided by national institutions, in particular the institutions participating in the {\it Gaia} Multilateral Agreement. 


\end{document}